\documentclass[runningheads]{llncs}

\usepackage[T1]{fontenc}
\def\doi#1{\href{https://doi.org/\detokenize{#1}}{\url{https://doi.org/\detokenize{#1}}}}
\usepackage{graphicx}
\usepackage{listings}
\usepackage{subcaption}
\usepackage{url}            
\usepackage{booktabs}       
\usepackage{amsfonts}       
\usepackage{nicefrac}       
\usepackage{microtype}      
\usepackage{multirow}
\usepackage{makecell}
\usepackage{hyperref}

\usepackage[normalem]{ulem}

\usepackage{xcolor}

\begin{document}
\title{Using the Order of Tomographic Slices as a Prior for Neural Networks Pre-Training}
\titlerunning{Using the Order of Tomographic Slices as a Prior for NN Pre-Training}
%
\author{Yaroslav Zharov\inst{1}
\thanks{The authors gratefully acknowledge the data storage service SDS@hd supported by the Ministry of Science, Research and the Arts Baden-Württemberg (MWK) and the German Research Foundation (DFG) through grant INST 35/1314-1 FUGG and INST 35/1503-1 FUGG.}
\and
 Alexey Ershov\inst{1} \and
 Tilo Baumbach\inst{1} \and
 Vincent Heuveline\inst{2}}

 \authorrunning{Y. Zharov et al.}

 \institute{Laboratory for Applications of Synchrotron Radiation (LAS), Karlsruhe Institute of Technology, Germany \\
  \email{\{yaroslav.zharov,ershov,tilo.baumbach\}@kit.edu} \\ \and
Engineering Mathematics and Computing Lab (EMCL),
 Interdisciplinary Center for Scientific Computing (IWR), Heidelberg University, Germany \\
\email{vincent.heuveline@uni-heidelberg.de}}

\maketitle              

\begin{abstract}
The technical advances in Computed Tomography (CT) allow to obtain immense amounts of 3D data.
For such datasets it is very costly and time-consuming to obtain the accurate 3D segmentation markup to train neural networks.
The annotation is typically done for a limited number of 2D slices, followed by an interpolation.
In this work, we propose a pre-training method \emph{SortingLoss}.
It performs pre-training on slices instead of volumes, so that a model could be fine-tuned on a sparse set of slices, without the interpolation step. 

Unlike general methods (e.g. SimCLR or Barlow Twins), the task specific methods (e.g. Transferable Visual Words) trade broad applicability for quality benefits by imposing stronger assumptions on the input data.

We propose a relatively mild assumption -- if we take several slices along some axis of a volume, structure of the sample presented on those slices, should give a strong clue to reconstruct the correct order of those slices along the axis.

Many biomedical datasets fulfill this requirement due to the specific anatomy of a sample and pre-defined alignment of the imaging setup. 
We examine the proposed method on two datasets: medical CT of lungs affected by COVID-19 disease, and high-resolution synchrotron-based full-body CT of model organisms (Medaka fish).
We show that the proposed method performs on par with SimCLR, while working 2x faster and requiring 1.5x less memory.
In addition, we present the benefits in terms of practical scenarios, especially the applicability to the pre-training of large models and the ability to localize samples within volumes in an unsupervised setup.

\end{abstract}

\begin{figure}[h]
\centering
\begin{subfigure}{.32\textwidth}
  \centering
  \includegraphics[width=1\linewidth]{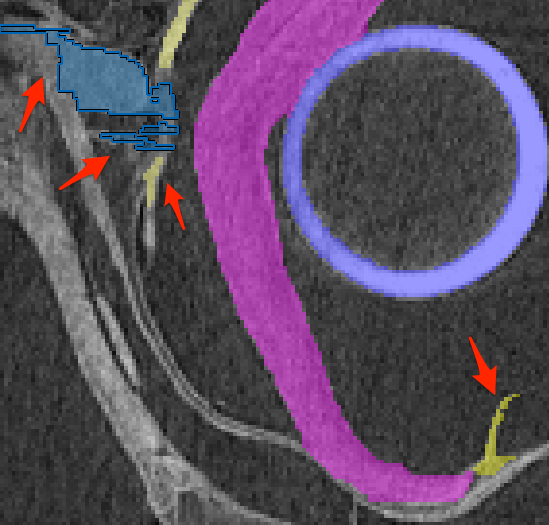}  
  \caption{expert segmentation}
  \label{fig:sub-first}
\end{subfigure}
\begin{subfigure}{.32\textwidth}
  \centering
  \includegraphics[width=1\linewidth]{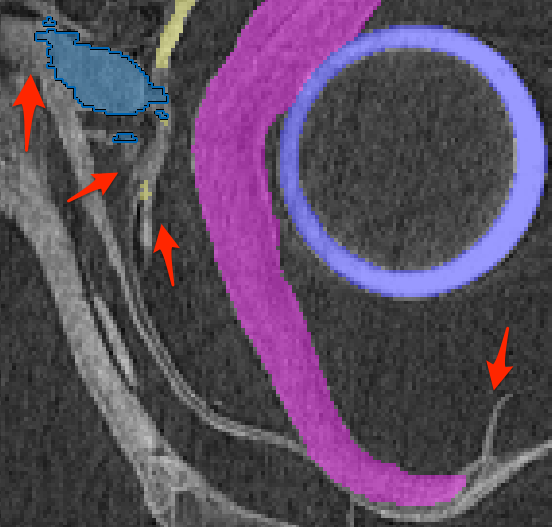}
  \caption{without pre-training}
  \label{fig:sub-second}
\end{subfigure}
\begin{subfigure}{.32\textwidth}
  \centering
  \includegraphics[width=1\linewidth]{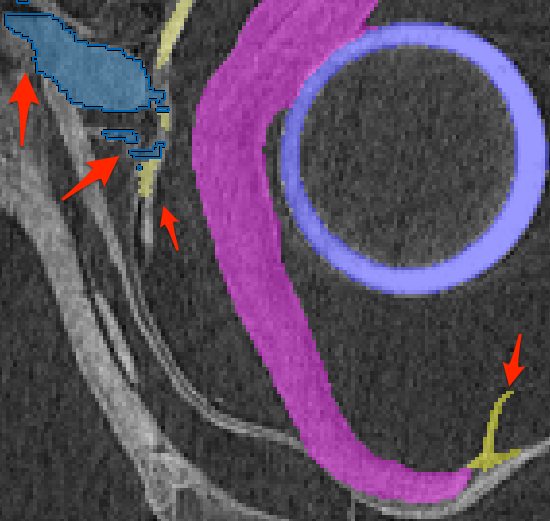}  
  \caption{with pre-training}
  \label{fig:sub-second}
\end{subfigure}
\caption{Comparison of fine structures of a visual system of Medaka fish segmented with and without the proposed pre-training method. The pre-trained network provides more accurate segmentation of smaller structures.}
\label{fig:fig}
\end{figure}



\section{Introduction}\label{sec:introduction}
Computed tomography (CT) is an important technique for medical diagnostics, materials research and other scientific applications.
Advances in imaging technologies and availability of CT facilities lead to  the widening gap between the capacity to produce huge amount of 3D data and the ability of domain experts to analyze it. 
This might be tackled using deep learning (DL) approaches, which are proven to provide good results for a wide variety of image analysis problems. 

Semantic segmentation is a typical task for further analysis, e.g. visualization of an object or quantitative measurements of morphological parameters. 
The annotation of massive amounts of CT data with pixel precision is an error-prone and tedious task.
The 3D ground truth markup is usually performed by domain experts on a limited number of 2D slices of the volume data and then interpolated.
This interpolation step could introduce errors (see example in Appendix \ref{app:interpolation}, Fig. \ref{fig:bad-interpolation}).

To exploit the full potential of large datasets, the self-supervised learning (SSL) might be employed.
Such methods aim to learn from the unlabeled data.  Usually, some proxy task is designed to perform the pre-training. Then, the model is transferred to a target task, making the final model more accurate or faster to train.
Since the rise of the SSL, the most attention was dedicated to the development of techniques for 2D photographic data.
However, tomographic data differs from 2D photographic images in several important aspects.
In photography, the projective geometry and varying viewpoint substantially change the appearance of a scene, objects and their relations. 
This leads, for example, to such problems as occlusions. 
On the other hand, CT imaging aims to reconstruct the undistorted 3D structure of an object. 
Moreover, imaging conditions and acquisition protocols are commonly fixed for a particular type of CT experiments. 
Also, the variability in appearance and positioning of background objects, such as sample holders, frames or tubes, in CT data is much smaller than in natural scenes.
This information can be exploited to devise a method, which relies on the specific 3D structure and features of an object of interest.

In this work we propose a pre-training method, which is intended to reduce the amount of manual markup and avoid interpolation errors by using only the expert-driven 2D slice markups instead of 3D labels.  

The paper is organized as follows. 
In \ref{sec:related-works} we describe related work and baseline methods. 
In \ref{sec:method} we describe the proposed method. 
Next, in \ref{sec:experiments} we outline our experiments' setup, and discuss comparisons with baselines and sensitivity experiments. 
We also discuss important practical questions related to the proposed method. 
Finally, we conclude on our contribution and possible further research.

\section{Related Work}\label{sec:related-works}

In the last years there is a growing interest in self-supervised learning (SSL). 
In this section we only give a short overview of the most popular baselines and methods that are closely related to our work.

A common approach for SSL is to design a human-interpretable proxy task to perform pre-training. 
One example is the work \cite{Noroozi2016} where the input image is cropped using a grid into a set of patches, which are then shuffled. 
A model learns to find a proper position for each crop.
Another example is the self-colorization task \cite{Zhang2016}.

Recently, a lot of research has focused on the so-called \textit{contrastive learning} \cite{Newell2020}. 
Contrastive SSL tasks are designed to distinguish between positive and negative examples. 
In \cite{Chen2020} authors presented SimCLR -- a method for contrastive learning, which quickly became a popular approach. 
It relies on a specific procedure of batch construction -- a number of samples are taken from a dataset, then two \emph{views} of each image are created by applying random augmentations. 
The model is trained to pull together embeddings of different views of the same image (positive samples) and push away from embeddings of different images (negative samples).
As a downside this approach requires a large batch size containing many negative samples. 
This leads to a higher memory footprint and longer computation time. Another difficulty for SimCLR is the requirement that the positive and the negative samples should be as diverse as possible. 
Currently, many methods aim to tackle this by imposing constraints on the embedded vectors themselves \cite{Bardes2021}. 
Moreover, it was noted that SimCLR might be sensitive to harder augmentations \cite{Wang2021}, requiring fine-tuning of the method. 
More details on various contrastive methods are given in the review \cite{Weng2021}.

Another approach for SSL is to use task-specific losses. These methods are based on the specific knowledge of an application area and dataset.
In the work by \cite{Spitzer2018} the authors propose to employ the geodesic distance between two patches of a human brain cortex.
Using geodesic distance was an important metric due to the inherent 3D structure of a brain cortex -- its characteristic folds with high curvature.
In another work \cite{Haghighi2021}, the authors proposed to rely on the fact that many medical images are highly structured (i.e. relative positions between organs) and aligned during a scanning procedure (e.g. X-ray image of lungs). 
Thus, they supposed that taking crops from the same location of an image will result in the same structures being demonstrated on the crop.
Authors therefore selected several pre-defined crop locations and assigned a separate class for each of those locations. 
These classes were called Transferable Visual Words. 
It is important to note, that the strong assumption about the alignment of data requires thorough data filtering, even for medical application.

In the next section, we will describe the proposed method and its key differences from the previous work.

\section{Proposed Method}\label{sec:method}
In our paper we aim to design a method specifically targeted at CT data, which removes the need for negative sampling by using a prior knowledge about data.
On the other hand, we aim to relax assumptions about samples to enable the method to work on a wider range of datasets.

In short, we require from our NN model to predict the correct order of slices taken from different positions within an individual volume. 
With respect to this, our method is similar to the method presented in \cite{Spitzer2018}. 
In contrast, we propose to estimate only the \emph{relative} positions instead of using the exact distances. 
In this way, we relax the assumption of size constancy of objects, imposed in \cite{Spitzer2018} to ensure that slicing is performed more or less at the same position of each sample, thus showing same structures. 
Like the method presented in \cite{Haghighi2021}, we expect similar structure (e.g. same organs and their arrangement) across different volumes.
Unlike it, we avoid the strict requirement of perfect alignment of \emph{all} volumes by comparing only the slices taken from a single volume.

We design our method \textit{SortingLoss} using the following assumptions: 

\begin{enumerate}
    \item The structure of a sample should be a strong clue for the ordering along at least one axis of the volume.
    Hereinafter, we call this axis the \emph{ordering axis}.
    For example, for a human body, the anatomy determines relative positions between a brain, a heart and kidneys, along the cranio-caudal axis.
    \item The ordering axes of all samples should be co-directed. 
    That is, for a human body, the ordering should be as follows: brain, heart and then kidneys (not vice versa). 
\end{enumerate}

We would like to stress, that other changes of composition are allowed -- rotation of a sample around the ordering axis, small tilts of a sample and overall size changes.

During the training, we form a batch by sampling random slices along the ordering axis.
In our experiments we sample slices with the uniform distribution of sampling probability. 
However, it is possible to adapt it based on a prior knowledge about the location of a sample or organ of interest within a volume. 
For the volumes with low resolution along the ordering axis, it's important to sample slices without replacement, to avoid limiting the amount of possible comparisons in one batch.

We add a linear layer on top of the model to convert an embedding vector to a floating-point number representing the predicted relative position.

We use the \emph{margin ranking} loss as it is described in \cite{Sculley2009}.
This loss enforces each pair with the known ranks to have the correct predicted order with a certain margin, not lower than a certain threshold, which we selected to be $0.2$ for all our experiments.
Other ranking losses could be beneficial to employ, since they could allow using the list-wise ordering information, instead of pairwise.
However, we let the question of benefits of such losses for a further investigation.

One benefit of SortingLoss is that during pre-training, we can track an interpretable metric instead of a plain loss. 
We propose a measure \emph{Mean Displacement}, which describes how much on average should a single slice change its position to match the perfect ordering:
$$MeanDisplacement = \frac{1}{bs} \sum_i \vert y_i - \mathtt{argsort}(f_{\theta}(\mathbf{x}))_i \vert$$

\section{Experiments}\label{sec:experiments}

To avoid competition of hyperparameters, we refrain from exhaustive hyper-parameters tuning.
Limited by the size of available GPU memory, we sample 16 distinct slices for each batch. Due to the use of paired samples in SimCLR, from the computational point of view, this corresponds to a batch size of 32.
Thus, our comparison favors the baseline.
For both pre-training and fine-tuning we use the same strong augmentations, which contains a color jitter, random rotations, noise (see Appendix \ref{app:augmentations}, Table \ref{tab:norm-augment}).

As proposed by \cite{Newell2020}, for comparisons we choose a simple training and fine-tuning approaches.
Where not stated otherwise, we perform experiments in the following manner.
First, we pre-train a ResNet-50 model for 70000 steps for each method using the same training data.
Then we produce 5 random train-test splits (75\% of volumes were used as a train set each time) and trained a DeepLabV3+ model with a pre-trained encoder for the amount of steps specified in each experiment. 
For both pre-training and fine-tuning, we use the Adam optimizer with the learning rate of $3e-4$.
As a test metric we calculate the mean Intersection over Union (IoU) on the test set.
For each experiment, we present the mean and standard deviation of those 5 runs.

To mimic a markup process in a low data regime we have chosen to limit the number of annotated slices, rather than limiting the number of training volumes.
Inspired by the findings presented in \cite{Zoph2020}, we compare the \emph{SortingLoss} both with the baseline method (SimCLR) and with training a model from scratch.

To evaluate our approach we use two datasets of different imaging objects and segmentation targets.

\emph{MosMedData dataset} presented in \cite{Morozov2020} contains 1110 volumes of the human chest CT images acquired by medical CT scanners.
Fifty volumes of this dataset contain segmentation of lung tissues affected by the COVID-19 disease.

\emph{Medaka visual system dataset} is a large-scale dataset for the ongoing project of the authors on genotype-phenotype mapping. We plan to open the dataset after the end of the project.
The dataset contains 340 high-resolution synchrotron-based $\mu$CT volumes of the Medaka fish (\textit{Oryzias latipes}) stained and imaged based on substantially improved protocols described in \cite{Weinhardt2018}.  
For 15 unique volumes we obtained 21 segmentation annotations with 5 classes of structures of a visual system (see Figure \ref{fig:fig}).
Due to having multiple markups for some of the volumes, we are able to estimate the quality of the segmentation by experts -- $74.8\%$ IoU, which shows that the dataset is challenging to annotate, even for the experts.

The description of data preparation for each dataset is listed in Appendix \ref{app:datasets-description}.

\subsection{Results}



\paragraph{MosMedData}
We performed pre-training on the entire MosMedData dataset and fine-tuning on 100\% and 10\% of annotated slices.
Since the performance on the 10\% was already low and 10\% subsample being equal to just 70 training slices, we did not take a smaller fraction.
From the results  presented in the Table \ref{tab:mosmed-covid} we conclude that the performance of SortingLoss is close to SimCLR.

\paragraph{Small-scale pre-training on Medaka dataset}
Since we have a limited computational capacity and cannot perform direct comparison with SimCLR on the full dataset, in this experiment we pre-trained a model on a dataset containing 50 random volumes.
We then performed fine-tuning on 100\%, 10\% and 1\% of annotated slices.
From the comparison presented in Table \ref{tab:medaka-small}, we conclude that SortingLoss always outperforms the baseline.

\paragraph{Large-scale pre-training on Medaka dataset}
We performed pre-training of the ResNet-152 model on random 250 volumes.
Since it took 13 days on 4xRTX2080Ti, we limited ourselves only to a comparisons with SortingLoss.
We found out, that even while fine-tuning U-Net (a model with a large decoder), and doing it in a full-data regime, we were able to improve IoU by 2.3\% compared to training from scratch.

\begin{table}[!htb]
    \begin{minipage}{.44\linewidth}
      \centering
        \begin{tabular}{r||c|c}
            & \multicolumn{2}{c}{slices \%}  \\
            pre-training & 10  & 100 \\ \hline \hline
            None &  $18.6 \pm 4.6$ & $34.4 \pm 7.5$ \\ \hline
            SimCLR & $\mathbf{21.8 \pm 7.7}$ & $\mathbf{40.1 \pm 3.9}$ \\ \hline
            SortingLoss & $20.8 \pm 5.7$ & $39.9 \pm 5.3$ \\ \hline
        \end{tabular}
        \caption{IoU for pre-training on the MosMedData dataset to segment tissues affected by COVID-19.}
        \label{tab:mosmed-covid}
    \end{minipage}%
    \hspace{.03\linewidth}
    \begin{minipage}{.53\linewidth}
      \centering
        \begin{tabular}{r||c|c|c}
            & \multicolumn{3}{c}{slices \%}  \\
            pre-training & 1 & 10 & 100 \\ \hline \hline
            None &  $57.4 \pm 3.4$ & $\mathbf{69.2 \pm 1.2}$ & $69.3 \pm 1.3$ \\ \hline
            SimCLR &  $58.5 \pm 3.9$ & $67.6 \pm 1.0$ & $67.9 \pm 2.0$ \\ \hline
            SortingLoss & $\mathbf{59.0 \pm 3.2}$ & $68.8 \pm 1.8$ & $\mathbf{69.9 \pm 0.9}$ \\ \hline
        \end{tabular}
        \caption{IoU for pre-training on the Medaka dataset to segment organs of a visual system.}
        \label{tab:medaka-small}
    \end{minipage}
\end{table}


\subsection{Sensitivity Study and Practitioner Questions}


\paragraph{Does harder augmentation yield better pre-training?}
As reported in \cite{Zoph2020}, the strong augmentation can diminish benefits of pre-training.
To check how our method performs under tougher augmentations, we increased magnitudes of all augmentations (see Appendix \ref{app:augmentations}, Table \ref{tab:harder-augment}).
From the results shown in Table \ref{tab:harder-aug} we can conclude, that our method is not only robust to tougher augmentation, but allows to improve the results and significantly outperform the baseline.


\begin{minipage}{\textwidth}
  \begin{minipage}[b]{0.44\linewidth}
    \centering
      \includegraphics[width=.95\textwidth,height=0.7\textheight,keepaspectratio]{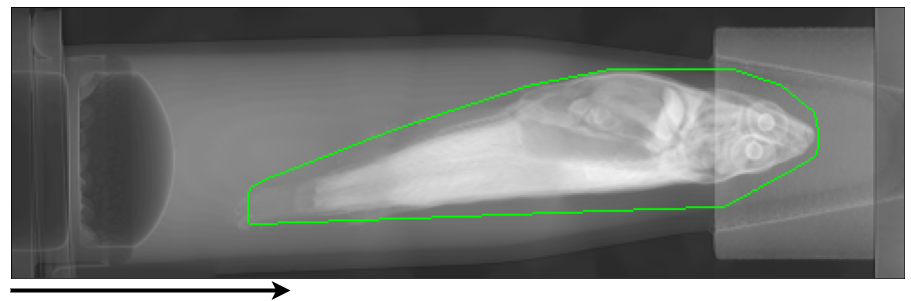}
      \captionof{figure}{Projection of a convex hull surrounding Medaka fish sample. The hull was built by estimating uncertainties of the local predictions of the model by injecting dropouts. Ordering axis is shown by an arrow.}
      \label{fig:fish-uncertainty-mask}
  \end{minipage}
  \hspace{0.02\linewidth}
  \begin{minipage}[b]{0.44\linewidth}
    \centering
    \begin{tabular}{r||c|c}
        &  \multicolumn{2}{c}{slices \%}  \\
        pre-training & 10 & 100 \\ \hline \hline
        SortingLoss & $\mathbf{+2.8}$ & $\mathbf{+2.4}$ \\ \hline
        SimCLR & $-3.9$ & $-2.9$ \\ \hline
    \end{tabular}
    \captionof{table}{Mean IoU difference after introducing tougher augmentations into the pre-training on the MosMedData dataset.}
    \label{tab:harder-aug}
    \end{minipage}
  \end{minipage}


\paragraph{Are there qualitative improvements visible to experts?}
\footnote{We would like to thank the Wittbrodt, Centanin and Weinhardt groups at the Centre for Organismal Studies at Heidelberg University, especially Tinatini Tavhelidse-Suck for providing access to the expert opinion.}
To evaluate the quality of segmentations from a human perspective, alongside with qualitative comparison presented in Fig. \ref{fig:fig}, we created a questionnaire where we have asked biology experts to select the best segmentation from three randomly presented slices of different markups. 
The first was for the neural network trained from scratch, the second for the model pre-trained using SimCLR and the third for the SortingLoss approach.
In total, we set 19 comparisons and questioned 11 experts.
We tested a hypothesis that the pre-trained models are better than the model trained from scratch with a one-sided paired t-test.
We can conclude that the SimCLR pre-training is clearly better than training from scratch (p-value = 0.02), while for SortingLoss the evidence is not strong enough (p-value = 0.14).
The comparison between two methods of pre-training does not yield definite result either (p-value = 0.16).

\paragraph{Is the sampling from the same volume beneficial for pre-training?}
As it is shown in work \cite{Mo2021}, contrastive methods can overrely on the background.
Since the background of different volumes may vary due to changes in imaging conditions, we hypothesized that sampling from the same volume, performed by SortingLoss, may be something beneficial on its own.
We performed pre-training using SimCLR on the full MosMedData dataset, while forming the whole batch only from the slices within a single volume.
For 10\% we've got $-2.8$ IoU points and for 100\% of slices $-0.8$.
We conclude, that the same-volume sampling, used for our method, did not lead to better results. 
Rather, it was an approach to get rid of the assumption on data alignment.


\paragraph{What are computational benefits?}
As we described in the beginning of the experimental section, during the comparisons, the baseline method has a favor -- its effective batch size is twice as large.
As a result, SimCLR requires more resources.
On a A5000 GPU card SortingLoss requires 40\% less memory (11 Gb vs 18.3 Gb) and trains two times faster (4.7 batches/second vs 2.3 batches/second).
To match computational resources, we trained SimCLR on the MosMedData dataset with two times less count of the images sampled to form a batch (therefore, same effective batch size).
As a result, for SimCLR the quality dropped by 3.3\% and became lower than the quality of SortingLoss.
This speed difference could be especially important for pre-training on large-scale datasets.
We have also compared the convergence speed.
The fine-tuning of a model pre-trained with any of the above mentioned methods converges to the final value of the non-pretrained model 2 times faster (measured in epochs).
Both pre-training methods do not differ in convergence speed.





\paragraph{Does the trained model really use the sample structures?}

We tested our main assumption -- that our loss will enforce the use of features of a sample.
We pre-trained a ResNet-18 model with our loss, but with additional dropouts added between the layer blocks.
After training, we used inference dropout to sample the distribution of the feature vectors of the model just before the final mean average pooling layer.
The Figure \ref{fig:fish-uncertainty-mask} shows the mask of the largest connected area of top 3\% of standard deviation of the vectors.
We tested the ability of this approach to localize a sample in an unsupervised manner. It was capable of removing 74\% of volume on average without cutting out anything other than tips of fins.

\section{Conclusion}\label{sec:conclusion}
In this paper we presented a novel method specifically tailored for pre-training on CT data. 
As we demonstrated with experiments on different datasets, it performs on par with the popular baseline -- SimCLR.
The proposed method utilizes prior knowledge about the inherent structure of a volumetric dataset. 
This allows us to remove the sampling of redundant image pairs and therefore improve the training speed and lower memory consumption.
As a pretext task, the method performs the ordering of 2D slices.
It allows fine-tuning of a pre-trained model on a sparse set of annotated slices without interpolation.
Other advantages of the method include: the positive response to harder augmentations, the ability to localize a sample without supervision, and the ability to improve results of the larger networks trained in the full-data regime.   
A limitation of the \emph{SortingLoss} method is a special assumption on the input data -- the structure of a sample should allow to perform the unique ordering of slices along at least one axis. 
However, many self-supervised pre-training techniques impose even stronger assumptions. 

The proposed method has a wide applicability.
The mild assumption on data could be met for many biomedical datasets, such as medical imaging of specific organs (e.g. brain, heart, pancreas), model organisms (e.g. fish, mice) or invertebrates (e.g. arthropods).
Despite that we have demonstrated the application of our method on X-ray CT data, it can also be applied to datasets from other tomographic imaging methods, such as MRI, PET, or neutron tomography.



\bibliographystyle{splncs04}
\bibliography{references}

\newpage
\appendix

\section{Segmentation Interpolation Errors}
\label{app:interpolation}
We present a visual comparison between the interpolation of the lung segmentation from the dataset presented in \cite{Ma2021} and segmentation model prediction.
The slice was taken along the sagittal plane.
\begin{figure}[h]
    \centering
  \includegraphics[width=0.3\linewidth]{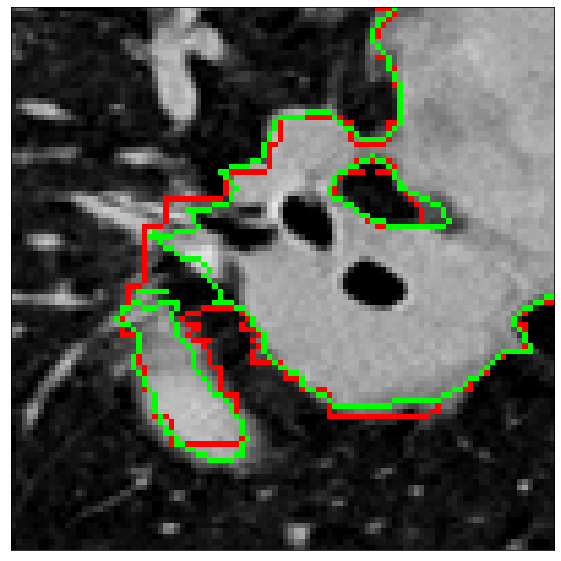}  
  \caption{Comparison of the expert interpolated markup (in red) with model prediction after training (in green).}
  \label{fig:bad-interpolation}
\end{figure}

\section{Used Augmentations}
\label{app:augmentations}

We used Albumentations library for the augmentations \cite{Buslaev2020}.
In tables \ref{tab:norm-augment} and \ref{tab:harder-augment}  we present a normal set of augmentations, which were used in all experiments for pre-training and fine-tuning, and tougher set of augmentations, used to evaluate the model response for tougher augmentation.

\begin{table}[]
    \centering
    \begin{tabular}{c|c|c}
        function & parameters & probabilities \\ \hline
        \hline
        \verb|Blur| & \verb|blur_limit=5| & \multirow{3}{*}{\texttt{0.5}} \\ \cline{1-2}
        \verb|MedianBlur| & \verb|blur_limit=5| \\ \cline{1-2}
        \verb|MotionBlur| & \verb|blur_limit=5| \\ \hline
        \verb|CLAHE| & -- & \multirow{2}{*}{\texttt{0.2}}\\ \cline{1-2}
        \verb|Equalize| & -- \\ \hline
        \verb|RandomBrightnessContrast| & \makecell{\texttt{brightness\_limit=0.2} \\ \texttt{contrast\_limit=0.2}} & \multirow{2}{*}{\texttt{0.2}}\\ \cline{1-2}
        \verb|RandomGamma| & \verb|gamma_limit=(90, 110)| \\ \hline
        \verb|GridDistortion| & \verb|distort_limit=0.2| & \multirow{3}{*}{\texttt{0.5}} \\ \cline{1-2}
        \verb|ShiftScaleRotate| & -- \\ \cline{1-2}
        \verb|GlassBlur| & \verb|max_delta=5| \\ \hline
        \verb|GaussNoise| & \verb|var_limit=(20, 50)| & \texttt{0.5}
    \end{tabular}
    \caption{Industry level augmentation used in all experiments for pre-training and fine-tuning. United cells with probability means that with given probability one random function applies.}
    \label{tab:norm-augment}
\end{table}

\begin{table}[]
    \centering
    \begin{tabular}{c|c|c}
        function & parameters & probabilities \\ \hline
        \hline
        \verb|Blur| & \verb|blur_limit=15| & \multirow{3}{*}{\texttt{0.5}} \\ \cline{1-2}
        \verb|MedianBlur| & \verb|blur_limit=15| \\ \cline{1-2}
        \verb|MotionBlur| & \verb|blur_limit=15| \\ \hline
        \verb|CLAHE| & -- & \multirow{2}{*}{\texttt{0.2}}\\ \cline{1-2}
        \verb|Equalize| & -- \\ \hline
        \verb|RandomBrightnessContrast| & \makecell{\texttt{brightness\_limit=0.4} \\ \texttt{contrast\_limit=0.4}} & \multirow{3}{*}{\texttt{0.2}}\\ \cline{1-2}
        \verb|RandomGamma| & \verb|gamma_limit=(30, 170)| \\ \cline{1-2}
        \verb|Solarize| & \verb|threshold=(64, 192)| \\ \hline
        \verb|GridDistortion| & \verb|distort_limit=0.2| & \multirow{3}{*}{\texttt{0.5}} \\ \cline{1-2}
        \verb|ShiftScaleRotate| & -- \\ \cline{1-2}
        \verb|GlassBlur| & \verb|max_delta=5| \\ \hline
        \verb|GaussNoise| & \verb|var_limit=(20, 50)| & \texttt{0.5}
    \end{tabular}
    \caption{Augmentation used to compare reaction for pre-training with tougher augmentation. United cells with probability means that with given probability one random function applies.}
    \label{tab:harder-augment}
\end{table}

\section{Data Preparation Pipelines}
\label{app:datasets-description}
In this section, we present the data preparation pipelines and technical characteristics of the datasets used in our experiments set.

\paragraph{MosMedData dataset}
 Volumes have lice size $512\times512$, and from $30$ to $35$ slices per volume. 
 To prepare this dataset, we have clipped al values with a range of Hounsfield units of $(-900; 500)$, rescaled this range to match $(0; 255)$, and saved the result as an 8bit data.

\paragraph{Medaka Eye Segmentation} 
Originally, each volume has slice size $2016\times2016$ pixels, and contains on average $6000$ slices. 
To reduce the memory footprint, we downsampled each volume 2 times for each axis. 
We also clipped the values range between $1$ and $99.95$ percentiles (taken for each volume separately), scaled values to match $(0; 255)$, and then converted those data to 8bit.


\end{document}